\documentclass[aps,prl,notitlepage,superscriptaddress,twocolumn]{revtex4-1}
\usepackage{graphicx,subfigure,epsfig}
 \usepackage{array,multirow}
\usepackage{dcolumn}
\usepackage{amssymb,amsmath,amsfonts,mathrsfs}
\usepackage{array}
\usepackage{times,setspace}
\usepackage{latexsym}
\usepackage{float,flafter,bm,bbm}
\usepackage{epstopdf,color,multirow}
\usepackage[colorlinks,linkcolor=blue,anchorcolor=blue,urlcolor=blue,citecolor=blue]{hyperref}
\usepackage{footnote}
\usepackage{booktabs}

\hypersetup{
    colorlinks=true,
    linkcolor=blue,
    filecolor=magenta,
    urlcolor=blue,
}

\begin{document}

\title {Pseudo-spin-polarized topological superconductivity in kagome RbV$_3$Sb$_5$}
\author{Xilin Feng}\thanks{These authors contributed equally to this work.}
\affiliation{Department of Physics, Hong Kong University of Science and Technology, Clear Water Bay, Hong Kong, China}
\author{Zi-Ting Sun}\thanks{These authors contributed equally to this work.}
\affiliation{Department of Physics, Hong Kong University of Science and Technology, Clear Water Bay, Hong Kong, China}
\author{Ben-Chuan Lin}\thanks{linbenchuan@iqasz.cn}
\affiliation{International Quantum Academy, and Shenzhen Branch, Hefei National Laboratory, Shenzhen,518048, China}
\author{K. T. Law}\thanks{phlaw@ust.hk}
\affiliation{Department of Physics, Hong Kong University of Science and Technology, Clear Water Bay, Hong Kong, China}
\date{\today}
\begin{abstract}
Kagome superconductors AV$_3$Sb$_5$ (A=K, Rb, Cs) have sparked considerable interest due to the presence of several intertwined symmetry-breaking phases within a single material. Interestingly, in a recent experiment, magnetic hysteresis was observed in the superconducting state through magnetoresistance measurements in RbV$_{3}$Sb$_{5}$ [Nature Comm \textbf{17}, 1310 (2026)], providing strong evidence of a spontaneous time-reversal symmetry breaking superconducting state. The magnetic hysteresis, combined with crystalline symmetry, imposes strong constraints on the possible pairing symmetries of the superconducting state. In this work, we propose that RbV$_3$Sb$_5$ is a nodal topological superconductor with pseudo-spin-polarized Cooper pairs. The pseudo-spin-polarized superconducting domains resemble the properties of ferromagnetic domains and induce hysteresis. Moreover, the nodal topological superconducting state possesses Majorana flat band modes at the sample boundary, which can be detected by tunneling experiments. 
\end{abstract}
\maketitle

\begin{figure}[t]
    \centering
    \includegraphics[width=0.5\textwidth]{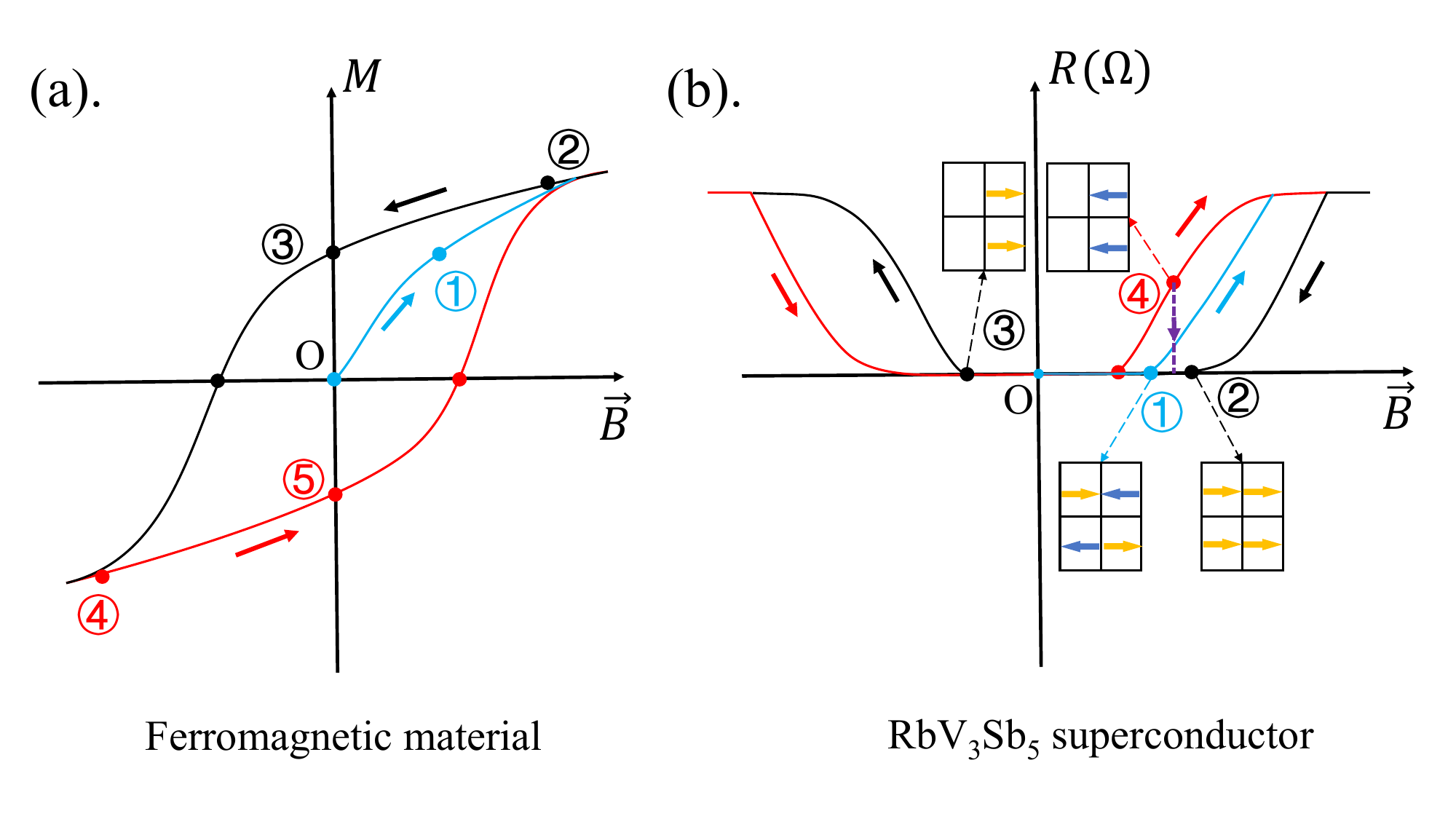}
    \caption{The hysteresis curves of a ferromagnet (a) and magnetoresistance of RbV$_3$Sb$_5$ (b), respectively. The initial magnetization (M) curve and the initial magnetoresistance (R) curve are shown in cyan, and the backward-sweep curve and forward-sweep curve are shown in black and red, respectively.  (b). The magnetoresistance of superconductor RbV$_3$Sb$_5$ as observed experimentally \cite{wang2024spinpolarized}. The insets illustrate the distribution of superconducting domains with different pseudo-spin polarizations at various steps along the initial resistance curve and the field sweeping process. Points 2 and 3 on the black curve indicate the critical magnetic fields at which the resistance starts to deviate from zero. The two points have different values and demonstrate the asymmetric behavior of the critical field during the field sweeping process. The dashed purple line represents the change in resistance after heating and recooling process at a fixed magnetic field as observed in the experiment. }
    \label{hyl}
\end{figure}

\emph{Introduction.}---The quasi-2D kagome superconductors AV$_3$Sb$_5$ (A=K, Rb, Cs) are widely studied due to the presence of several intriguing phases in this family of materials \cite{ortiz2019new,ortiz2020cs,jiang2023kagome,yin2022,wang2023,guguchia2023}, such as the chiral flux phase ~\cite{feng2021chiral,denner2021,lin2021complex,park2021electronic,feng2021low,christensen2021,jiang2021,yu2021,yang2020,yu2021conc}, the charge bond order state~\cite{wagner2023,wu2022charge,wang2023structure}, the nematic phase \cite{xu2022three,li2022rotation,grandi2023,tazai2023charge,jiang2023observation,Grandi2024}, the chiral excitonic ordered state~\cite{scammell2023chiral,ingham2025vestigial}, the superconducting state~\cite{wu2021nature,ortiz2021superconductivity,luo2021,yin2021superconductivity,ni2021anisotropic,zhao2021cascade,zhao2021nodal,wang2021charge,Astrid2022}, and the pair density wave (PDW) state~\cite{wu2023pair,jin2022interplay,zhou2022chern,schwemmer2023pair,deng2024chiral,yan2024chiral}. However, despite extensive investigations into the superconducting states of kagome superconductors AV$_3$Sb$_5$ (A = K, Rb, Cs), the nature of the superconducting states in these materials remains debated~\cite{mu2021s,holbaek2023unconventional,le2024superconducting,dai2024existence,hossain2025unconventional}.
Recently, the in-plane magnetic response of the superconducting state in RbV$_3$Sb$_5$ was found to exhibit highly nontrivial behavior, as reported in Ref.~\cite{wang2024spinpolarized}. In contrast to CsV$_3$Sb$_5$ samples measured by the same group, where the superconducting state shows no magnetic hysteresis, RbV$_3$Sb$_5$ exhibits magnetoresistance hysteresis under an in-plane magnetic field ~\cite{wang2024spinpolarized}. The hysteresis curves of RbV$_3$Sb$_5$ are shown in Fig.\ref{hyl} (b), resembling the hysteresis curves of ferromagnets. Interestingly, at a finite resistance state (Point 4 of the forward sweeping curve, for example), heating the sample above the critical superconducting temperature and re-cooling it drives the sample to zero resistance.

The unconventional magnetic responses of RbV$_3$Sb$_5$ strongly suggest the intrinsic time-reversal symmetry breaking nature of the superconducting state. The question is, what is the nature of this time-reversal symmetry breaking superconducting state? Considering the novel magnetic responses, the nematic normal state ~\cite{xu2022three}, and the nodal pairing structure \cite{guguchia2023tunable}, we propose that the superconducting state in RbV$_3$Sb$_5$ is an odd-parity, pseudo-spin-polarized (i.e., nonunitary) nodal topological state. 

\begin{figure}
    \centering
    \includegraphics[width=1\linewidth]{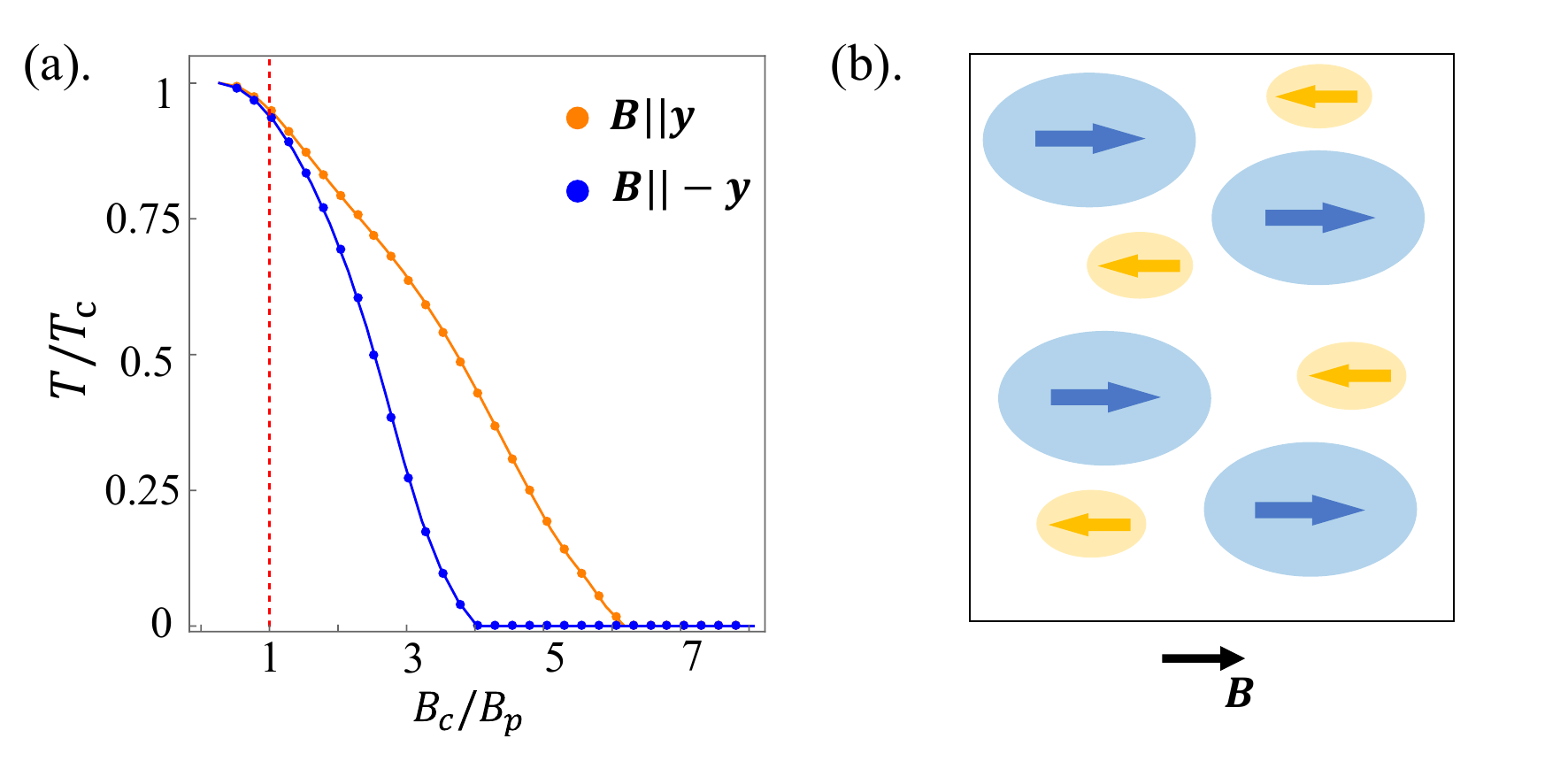}
    \caption{
    (a). Upper critical fields ($B_c$) calculated from the gap equation with $\phi = 0.3\pi$. The orange and blue lines show the upper critical field as a function of temperature for magnetic fields applied along the $+y$ and $-y$ directions, respectively. Here, the pseudo-spin polarization of the superconducting state is pinned to the $y$ direction, resulting in a higher upper critical field for a parallel field than for an anti-parallel one. The red dashed line represents the Pauli limit ($B_p$) estimated using the zero-field transition temperature. (b). A schematic picture of the superconducting domains when the magnetic field increases beyond Point 1 in the initial curve (blue line in Fig.\ref{hyl}(b)). The size of the superconducting domain with pseudo-spin polarization parallel to the applied magnetic field is larger than that with pseudo-spin polarization antiparallel to the field.
    }
    \label{H_Tc}
\end{figure}

In this work: 1) first of all, we construct an eight-band model for the normal state of RbV$_3$Sb$_5$ that incorporates the nematic state~\cite{xu2022three} and spin-orbit parity coupling (SOPC), and then define the pseudo-spin basis by projecting this model onto the two degenerate bands near the Fermi surface. 2) We suggest that the d-vector of the pseudo-spin-polarized superconducting state is a mixture of basis functions from different irreducible representations of the $D_{2h}$ point group. Although the pseudo-spin of Cooper pairs is polarized, there exists a unitary component in the real spin basis, allowing the pairing to be suppressed by a Zeeman field. 3) Superconducting domains with  pseudo-spin polarization parallel or anti-parallel to the applied magnetic field have different upper critical fields, as shown in Fig.\ref{H_Tc} (a). The novel magnetic hysteresis, as shown in Fig.\ref{hyl}(b), can be explained by the presence of superconducting domains with different pseudo-spin polarizations. 4) The pseudo-spin-polarized state is a nodal topological superconducting state with Majorana flat band modes at the sample edges. These Majorana modes can induce sharp tunneling conductance peaks in tunneling experiments. Overall, we suggest that RbV$_3$Sb$_5$ is the long-sought-after time-reversal symmetry breaking topological superconductor.

\begin{figure}[t]
    \centering
    \includegraphics[width=0.45\textwidth]{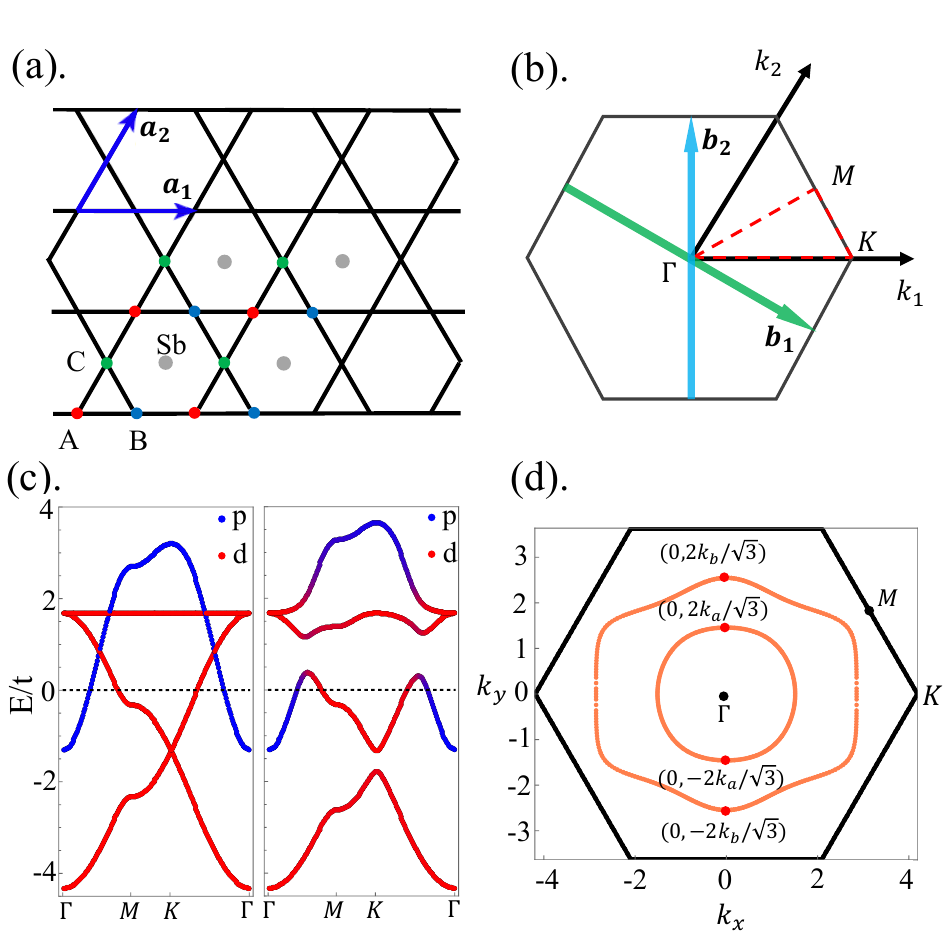}
    \caption{(a). The real space structure of the kagome lattice plane in RbV$_{3}$Sb$_{5}$. The arrows $\bm{a}_{1}$ and $\bm{a}_{2}$ are the unit vectors in the real space coordinate. The red, blue, and green atoms represent A, B, and C sublattices of V atoms, respectively. The grey atoms represent the in-plane Sb atoms. (b). The Brillouin zone of the kagome lattice. The green arrow $\mathbf{b}_{1}$ and cyan arrow $\mathbf{b}_{2}$ are the reciprocal lattice vectors. The red dashed line represents the high symmetry line in the Brillouin zone. The $k_{1}$ and $k_2$ show the coordinate in the reciprocal lattice. (c). Band structure along the high-symmetry line of the Brillouin zone without (left) and with (right) SOPC terms. The SOPC terms gives rise to two degenerate bands at the Fermi energy. In (c), to clearly illustrate the effect of SOPC, we omit the nematic normal state by setting $\zeta = 1$. (d). The Fermi surfaces (orange lines) of the eight-band model, considering the effects of both SOPC and a nematic normal state ($\zeta=2$). The red dots on the Fermi surfaces denote the four point nodes of the superconducting state described by $H_{BdG}(\mathbf{k})$.}
    \label{stru}
\end{figure}

\emph{The normal state Hamiltonian for RbV$_{3}$Sb$_{5}$.}--- First of all, we construct an eight-band model for the normal state of kagome RbV$_{3}$Sb$_{5}$. Fig.~\ref{stru}(a) shows the real space structure of the kagome plane in RbV$_{3}$Sb$_{5}$. The $d$-orbitals of V atoms and the $p_{z}$-orbitals of the in-plane Sb atoms have different parities, giving rise to the SOPC terms between them~\cite{xie2020spin,Gu2022,zhang2023spin}.  This term gives rise to two degenerate isolated bands [See Fig.~\ref{stru}(c)] near the Fermi energy and mixes the spin and momentum degrees of freedom, even in the presence of inversion.

The normal state Hamiltonian can be expressed as:
\begin{equation}
    H=\sum_{k, s, s^{\prime}} \Psi_{\mathbf{k},s}^{\dagger} H^{8-band}_{\mathbf{k},ss^{\prime}} \Psi_{\mathbf{k},s^{\prime}},
\end{equation}
where $s$ and $s'$ represent the spin index, and the specific form of $H^{8-band}_{\mathbf{k},ss^{\prime}}$ can be found in the \textbf{End Matter}.
The basis of the Hamiltonian is $\Psi_{\mathbf{k},s}=(c_{\mathbf{k},V_{A},s},c_{\mathbf{k},V_{B},s},c_{\mathbf{k},V_{C},s},c_{\mathbf{k},Sb,s})^{T}$, where $V_{A,B,C}$ denotes the sublattice sites of the $V$ atoms, as shown in Fig.\ref{stru}(a). The Brillouin zone and high symmetry lines of the kagome lattice are shown in Fig.~\ref{stru}(b), and
the band structures without and with SOPC along the high symmetry line are shown on the left and right sides of Fig.~\ref{stru}(c), respectively. It is worth noting that although the time-reversal symmetry breaking (TRSB) chiral flux phase exists in the normal states of the kagome superconductors AV$_3$Sb$_5$~\cite{feng2021chiral,denner2021,lin2021complex,park2021electronic,feng2021low,christensen2021,jiang2021,yu2021,yang2020,yu2021conc}, this type of TRSB originates from loop currents that generate out-of-plane orbital magnetization and is thus irrelevant to the hysteresis observed under in-plane magnetic fields in the superconducting state. Therefore, we do not include the effects of TRSB chiral flux phase in this work.

\emph{The construction of the pseudo-spin basis.}---In the presence of SOPC, spin is no longer a good quantum number. However, due to the presence of inversion and time-reversal symmetry, the bands of the 8-band model $H^{8-band}_{\mathbf{k},ss^{\prime}}$ remain doubly degenerate at every $\mathbf{k}$ point. Therefore, we can define a pseudo-spin basis in which the pseudo-spin transforms identically to real spins under symmetry operations. This pseudo-spin basis is referred to as the manifestly covariant Bloch basis (MCBB) \cite{yip2013model,fu2015parity,yip2016pseudospin,fischer2023superconductivity}.
To construct the MCBB, we project the eight-band model onto the two selected degenerate bands near the Fermi energy. Then the pseudo-spin operators $\tilde{\sigma}_{i}(\mathbf{k}),~(i=x,y,z)$ are derived by applying the projector onto the spin operators $\sigma_{i} \otimes I_{4\times 4}~(i=x,y,z)$ in the real spin basis. Subsequently, we can expand these pseudo-spin operators using linear combinations of the Pauli matrices $\rho_{i}\,(i=x,y,z)$. (Further details can be found in \textbf{End Matter}.)

When applying a magnetic field $\mathbf{B}=(B_{x},B_{y},B_{z})$, the effective two-band Hamiltonian can be written as follows:
\begin{equation}\label{effsingb}
    H_{0}(\mathbf{k})=\sum_{\mathbf{k} \alpha}\xi_{\mathbf{k}} c_{\mathbf{k}\alpha}^{\dagger}c_{\mathbf{k}\alpha}-\mu_{B}\sum_{\mathbf{k},\alpha,\beta}c_{\mathbf{k}\alpha}^{\dagger}B_{i}a_{ij}(\mathbf{k})\rho_{j}^{\alpha \beta}c_{\mathbf{k}\beta},
\end{equation}
where $\xi_{\mathbf{k}}$ represents the band dispersion near the Fermi energy, and the coefficients $a_{ij}(\mathbf{k})$ arise from expanding the pseudo-spin operators using Pauli matrices. In Eq.~(\ref{effsingb}), we define an effective magnetic field: $g_{j}(\mathbf{k})=-\mu_{B}\sum_{i}B_{i}a_{ij}(\mathbf{k})$, which acts on the pseudo-spin basis.

\emph{The pairing symmetry and superconducting order parameter.}--- Considering the effect of nematicity in the normal state~\cite{xu2022three}, the normal state point group symmetry of RbV$_3$Sb$_5$ becomes $D_{2h}$. As suggested by the magnetoresistance hysteresis in Fig.\ref{hyl} (b), we expect the Cooper pairs to couple to the Zeeman field. Therefore, within the MCBB, we look for pseudo-spin triplet pairing states within the odd-parity irreducible representations (irreps) of $D_{2h}$. The order parameters belonging to the odd parity irreps are listed in Table.\ref{REP1main}.
\begin{table}[t]
     \caption{The irreducible representations (irreps) of $D_{2h}$ point group and the corresponding order parameters. For our purposes, only the Irreps with odd parity are listed. In this table, $k_{1}=k_{x}$, $k_2=k_x/2+\sqrt{3}k_y/2$, and $k_{3}=k_{2}-k_{1}$. In addition, $\mathbf{\hat{x}}$, $\mathbf{\hat{y}}$, and $\mathbf{\hat{z}}$ represent axial-vectors oriented along the $x$, $y$, and $z$ directions, respectively.}
    \centering
    \begin{ruledtabular}
    \begin{tabular}{c|cc}\label{REP}
    Representations & order parameter $\mathbf{d}(\mathbf{k})$ \\ \hline
    $A_{u}$ & $\sin(k_{1})\mathbf{\hat{x}}$; $(\sin(k_{2})+\sin(k_{3}))\mathbf{\hat{y}}$.\\ \hline
    $B_{1u}$ & $(\sin(k_{2})+\sin(k_{3}))\mathbf{\hat{x}}$; $\sin(k_{1})\mathbf{\hat{y}}$. \\
    \hline
    $B_{2u}$ & $(\sin(k_{2})+\sin(k_{3}))\mathbf{\hat{z}}$\\ \hline
    $B_{3u}$ & $\sin(k_{1})\mathbf{\hat{z}}$ \\
    \end{tabular}
    \end{ruledtabular}
    \label{REP1main}
\end{table}
If the superconducting state belongs to a single irrep, it must be unitary and thus cannot account for the observed hysteresis.
Therefore, a natural consideration to construct the nonunitary pairing is to combine different irreps, assuming that they have nearly degenerate critical temperatures~\cite{jiao2020chiral,hayes2021multicomponent,shaffer2022chiral,aoki2022unconventional,ishihara2023chiral}. 
By incorporating the nodal properties of the superconducting states, as suggested in Ref.\cite{guguchia2023tunable}, we choose, without loss of generality, the order parameters with point nodes located along the $k_{1}=0$ axis. The nonunitary order parameter with an in-plane magnetic moment can generally be written as:
\begin{equation}\label{dvect}
    \mathbf{d}(\textbf{k})=\Delta_0 \sin(k_{1})(\cos(\theta)\cos(\phi),\sin(\theta)\cos(\phi),i\sin(\phi)),
\end{equation}
where the parameter $\theta$ controls the direction of the pseudo-spin polarization of the Cooper pairs in MCBB. For example, when $\theta =0$, the order parameter $\mathbf{d}(\textbf{k})=\mathbf{d}_{A_{u}+i B_{3u}}(\textbf{k})=\sin(k_{1})(\cos(\phi),0,i\sin(\phi))$ represents a nodal superconducting state with pseudo-spin polarization aligned with the $y$-direction. The parameter $\phi$ controls the amplitude of the Cooper pair's pseudo-spin polarization, which is linearly proportional to $\eta= \sin(2\phi)$, with $\phi$ ranging from $0$ to $\pi/2$.
For $\phi=0$ or $\phi=\pi/2$, the amplitude of the Cooper pair's pseudo-spin polarization is zero, resulting in a unitary superconducting state. In contrast, for $\phi=\pi/4$, the superconducting state is non-unitary and fully polarized, with half of the electrons remaining unpaired. 

\emph{The upper critical field.}---In this section, within the MCBB, we calculate the upper critical field of superconductivity in the monolayer limit by using the linearized gap equation. The gap equation can be written as \cite{frigeri2004superconductivity,sigrist2009introduction}:
\begin{equation}\label{gapeq}
\begin{aligned}
        \Delta_{ss'}(\mathbf{k})=&-k_{B}T\frac{1}{N_{k}}\sum_{n,\mathbf{k}'}\sum_{s_{1},s_{2},s_{3},s_{4}}V_{\mathbf{k},\mathbf{k}';s,s',s_{1}s_{2}}\\ &G^{0}_{s_{1}s_{3}}(\mathbf{k}',i\omega_{n})\Delta_{s_{3},s_{4}}(\mathbf{k}')G^{0}_{s_{4}s_{2}}(-\mathbf{k}',-i\omega_{n})^{T},
\end{aligned}
\end{equation}
where $G^{0}_{s,s'}(\mathbf{k},i\omega_{n})$ is the Matsubara Green‘s function constructed by the effective two-band Hamiltonian in Eq.(\ref{effsingb}). 

We simplify the gap equation by using the decoupling of the interaction \cite{sigrist2009introduction}:
\begin{equation}\label{inter}
     V_{\mathbf{k},\mathbf{k}';s,s',s_{1},s_{2}}=v (i \mathbf{d}(\textbf{k})\cdot \bm{\rho}\rho_{2})_{s,s'}(i\mathbf{d}(\mathbf{k}')\cdot \bm{\rho} \rho_{2})^{\dagger}_{s_{1}s_{2}}.
\end{equation}
The decoupling constant is represented by $v$, which is proportional to the strength of the interaction. 
Furthermore, in the experiment, the magnetoresistance hysteresis appears even with zero field cooling, and the zero field superconducting diode effect was observed \cite{wang2024spinpolarized}. These experiments strongly suggest the breaking of time-reversal symmetry without an external field. 
We therefore choose an intrinsic pseudo-spin polarization with amplitude $\eta_{0}=\sin(2\phi_{0})$, where  $\phi_{0}=0.3\pi$ is used in Fig.\ref{H_Tc}(a). An appropriate decoupling constant $v$ is chosen accordingly to ensure that the $\phi=\phi_{0}$ state has the lowest free energy at zero Zeeman field. With the given parameters, we can solve the gap equation at a finite field to obtain the upper critical field of the $\phi=0.3\pi$ state when the magnetic field is parallel or anti-parallel to the pseudo-spin polarization direction. The results of the upper-critical fields are shown in Fig.\ref{H_Tc}(a), which demonstrates that for a superconducting domain with pinned pseudo-spin polarization, the upper-critical fields are different for magnetic fields parallel or anti-parallel to the pseudo-spin polarization direction. Therefore, we expect that at a given temperature, superconducting domains with pseudo-spin polarization anti-parallel to the Zeeman field will have a higher energy and can shrink in size as the Zeeman field increases. This point is important to understand the magnetoresistance hysteresis, as explained in the next section. 

\emph{The origin of the hysteresis.}---To give a possible explanation for the magnetoresistance hysteresis observed in the experiment and depicted in Fig.\ref{hyl}(b), we first assume that there is an equal mixing of superconducting domains with pseudo-spin polarization pointing to the right (R-domains) and domains pointing to the left (L-domains). As the magnetic field, which points to the right, increases from zero [the initial curve denoted as cyan in Fig.\ref{hyl}(b)], the L-domains shrink in size as their pseudo-spin polarization is opposite to the applied Zeeman field. The shrinking of domains results in a finite resistance state as the magnetic field increases beyond Point 1. A schematic picture of the superconducting domains beyond Point 1 is shown in Fig.\ref{H_Tc}(b). During the backward sweeping process (the black curve), at Point 2, the superconducting domains are all R-domains, and the sample has zero resistance. When the magnetic field changes sign, R-domains start to shrink. Beyond Point 3, as the R-domains shrink further, zero resistance paths cannot be formed, and the sample goes to a finite resistance state.  Even though the actual domain dynamics can be much more complicated, this simple domain shrinking picture is consistent with the experimental observations. Moreover, the domain shrinking picture is also consistent with the fact that the critical field at Point 1 has a larger magnitude than the critical field at Point 3. This is because, at Point 3, more domains have pseudo-spin polarizations that are anti-parallel to the magnetic field, so the percolating superconducting paths are suppressed more rapidly by the Zeeman field than at Point 1.

Importantly, our superconducting domain picture is consistent with another novel observation of the hysteresis curve  \cite{wang2024spinpolarized}. For example, in Fig.\ref{hyl}(b), at Point 4 of the forward sweeping curve with finite resistance, applying a large current exceeding the critical current heats up the sample and destroys superconductivity. When the sample is re-cooled at a fixed magnetic field, it goes to the zero resistance state. This clearly shows that Point 4 is a metastable state, with the pseudo-spin polarization of the superconducting domains antiparallel to the applied magnetic field. Heating the sample at a fixed magnetic field destroys all domains. Upon recooling, superconducting domains with pseudo-spin polarizations parallel to the applied field are created with a larger size. This leads to a zero-resistance state due to the formation of percolating superconducting paths.

\emph{Topological properties and experimental detection.}--- In this section, we explore the topological properties of the superconducting state with the pairing d-vector proposed in Eq.(\ref{dvect}) with $\theta = 0 $ and $\phi = 0.3\pi $. The corresponding Bogoliubov-de Gennes (BdG) Hamiltonian at zero-field is:
\begin{equation}\label{bdgzfc}
\begin{aligned}
     H_{BdG}(\mathbf{k})=\left(\begin{array}{cc}H_{0}(\mathbf{k})&i\textbf{d}(\textbf{k})\cdot\bm{\rho}\rho_{y}\\(i\textbf{d}(\textbf{k})\cdot\bm{\rho}\rho_{y})^{\dagger}&-H_{0}^{*}(-\mathbf{k})\end{array}\right),
\end{aligned}
\end{equation}
where $\rho_{i} (i=x,y,z)$ represents the Pauli matrix in the pseudo-spin basis, and $\bm{\rho}=(\rho_{x},\rho_{y},\rho_{z})$. In $H_{0}(\mathbf{k})$, which is given by Eq.(\ref{effsingb}), the magnetic field $B_{i}$ is taken to be $0$. 

\begin{figure}[t]
    \centering
    \includegraphics[width=0.5\textwidth]{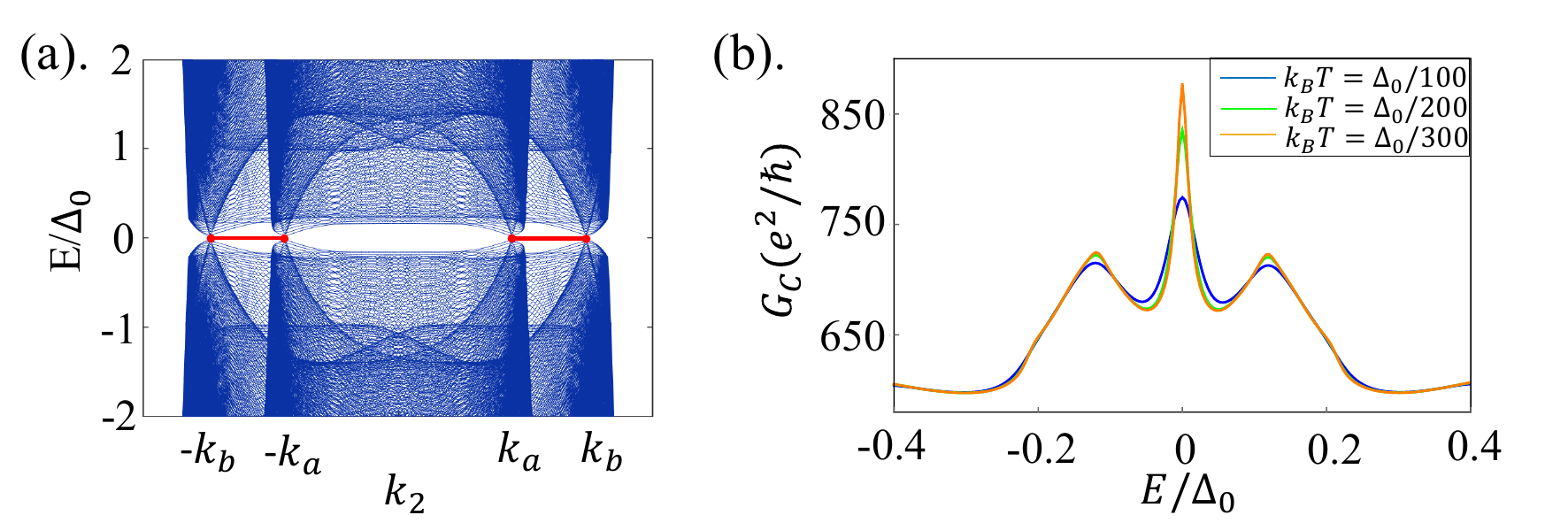}
    \caption{ Majorana flat band and edge tunneling conductance in RbV$_3$Sb$_5$: (a). The energy spectrum of $H_{BdG}(\mathbf{k})$ with open boundary condition in the $k_1$ direction and periodic boundary condition in the $k_2$ direction. Zero energy Majorana modes appear when $-k_b < k_2 < -k_a$ and $k_a < k_2 < k_b$. The amplitude of the gap is chosen to be $\Delta_0=0.03 t$. (b). The edge tunneling conductance at different temperatures $k_BT=\Delta_0/100,\Delta_0/200,\Delta_0/300$. The zero-bias peak is robust at a wide range of temperatures.
    }
    \label{open_tun}
\end{figure}

In the normal state, there are two Fermi surfaces enclosing the $\Gamma$ point of the Brillouin zone, as shown in Fig.\ref{stru}(d). Additionally, the zeros of the order parameter $\mathbf{d}$ lie along the $k_{1}=0$ axis. Consequently, the superconducting state described by $H_{BdG}(\mathbf{k})$ has four point nodes. The $k_{2}$-coordinates of the four point nodes are denoted as $\pm k_{a}$ and $\pm k_{b}$, respectively, as shown in Fig.\ref{stru}(d). The energy spectrum with open boundary conditions along the $k_x = k_1$ direction and periodic in the $k_2 = \frac{1}{2}k_x+\frac{\sqrt{3}}{2}k_{y}$ direction is shown in Fig.\ref{open_tun}(a). It is clear that there are Majorana flat bands joining the projections of the nodal points in the $k_2$ direction
\cite{sato2009topological,hasan2010colloquium,sau2010generic,alicea2010majorana,oreg2010helical,schnyder2011topological,qi2011topological,alicea2012new,beenakker2013search,wong2013majorana,schnyder2015topological,zhou2016ising}. It is verified that for fixed $k_2$,  the effective one dimensional Hamiltonian $H_{BdG}(\mathbf{k}) $ belongs to the BDI class, and Majorana modes appear when $-k_b < k_2 < -k_a$ and $k_a < k_2 < k_b$, as shown in the Supplemental Material \cite{supp}.

Due to the presence of a large number of zero energy Majorana modes, we expect that large zero bias conductance peaks can appear in tunneling experiments in RbV$_3$Sb$_5$ when the tunneling leads are coupled to appropriate edges. In the zero-field case, we calculate the tunneling spectra of the model in Eq.~(\ref{bdgzfc}) using the recursive Green's function method \cite{fisher1981relation,lee1981anderson,sun2009quantum,liu2012zero} (The details for the calculation can be found in the Supplemental Material \cite{supp}). As shown in Fig.~\ref{open_tun}(b),  a zero-bias conductance peak emerges as a consequence of the Majorana induced resonant Andreev reflections \cite{law2009majorana,wimmer2011quantum}. Therefore, the observation of the tunneling peaks can serve as evidence of nodal topological superconductivity.

\emph{Conclusion.}---In conclusion, RbV$_3$Sb$_5$ is a pseudo-spin-polarized superconductor with nontrivial topological properties. We establish a faithful eight-band model that incorporates the nematic normal state and spin-orbit parity coupling effects. Using this model, we calculate the upper critical fields for superconducting states with pseudo-spin polarizations parallel and antiparallel to the applied field. The difference between these two upper critical fields, combined with the pseudo-spin polarized superconducting domain picture, explains the observed hysteresis. We further suggest that the superconducting state in RbV$_3$Sb$_5$ is an odd-parity, pseudo-spin-polarized (i.e., nonunitary) nodal topological state with Majorana flat band edge modes.

\emph{Acknowledgements}---K. T. L. acknowledges the support of the Ministry of Science and Technology, China, The New Cornerstone Foundation, The State Key Laboratory of Quantum Information Technologies and Materials at CUHK, and the Hong Kong Research Grants Council through Grants No. MOST23SC01-A, No. RFS2021-6S03, No. C6053-23G, No. AoE/P-701/20, AoE/P-604/25R, No. 16309223, No. 16311424 and No. 16300325.
\bibliographystyle{apsrev4-2}
\bibliography{ref}

\section{End matter}
\emph{The model Hamiltonian of RV$_3$Sb$_5$.}---Under the basis $\Psi_{\mathbf{k},s}=(c_{\mathbf{k},V_{A},s},c_{\mathbf{k},V_{B},s},c_{\mathbf{k},V_{C},s},c_{\mathbf{k},Sb,s})^{T}$, the equal-spin part of the Hamiltonian is $H_{\sigma \sigma}(\boldsymbol{k})=\text{diag}\left(H_{V}(\boldsymbol{k}),H_{S b}(\boldsymbol{k})\right)$, in which
\begin{widetext}
\begin{equation}
\begin{aligned}
    H_{V}(\boldsymbol{k})=\left[\begin{array}{ccc}
-\mu_V & -2 t_1^{'} \cos \left(k_1 / 2\right) & -2 t_1 \cos \left(k_2 / 2\right) \\
-2 t_1^{'} \cos \left(k_1 / 2\right) & -\mu_V & -2 t_1 \cos \left(k_3 / 2\right) \\
-2 t_1 \cos \left(k_2 / 2\right) & -2 t_1 \cos \left(k_3 / 2\right) & -\mu_V
\end{array}\right],
\end{aligned}
\end{equation}
\begin{equation}
H_{S b}(\boldsymbol{k})=-2 t_2\left[\cos \left(k_1\right)+\cos \left(k_2\right)+\cos \left(k_3\right)\right]-\mu_{S b},
\end{equation}
where $k_1=k_x$, $k_2=k_x / 2+\sqrt{3} k_y / 2$, $k_3=-k_x / 2+\sqrt{3} k_y / 2$, $t_1^{'}=t_{1}*\zeta$, $\zeta$ show the nematic properties of the normal states. $t_{1}$ is the hopping strength between the nearest neighbor $V$ atoms, $t_{2}$ is the hopping strength between the nearest neighbor $Sb$ atoms, and $\mu_{V}$ and $\mu_{Sb}$ are the on-site energies of the $V$ and $Sb$ atoms, respectively. The spin-orbit-parity coupling (SOPC) part is expressed as:
    \begin{equation}
H_{\uparrow \downarrow}(\boldsymbol{k})=\left[\begin{array}{cccc}
0 & 0 & 0 & b \sin \left(\frac{k_3}{2}\right)(-\sqrt{3}-i) \\
0 & 0 & 0 & b \sin \left(\frac{k_2}{2}\right)(-\sqrt{3}+i) \\
0 & 0 & 0 & 2 b \sin \left(\frac{k_1}{2}\right) i \\
b \sin \left(\frac{k_3}{2}\right)(-\sqrt{3}-i) & b \sin \left(\frac{k_2}{2}\right)(-\sqrt{3}+i) & 2 b \sin \left(\frac{k_1}{2}\right) i & 0
\end{array}\right]
\end{equation}
and $H_{\downarrow \uparrow }(\mathbf{k})=H_{\uparrow \downarrow}^{\dagger}(\mathbf{k})$. $b$ represents the strength of the SOPC terms. In the main text, for the eight-band model, we set $t = 1$, $t_2 = 0.5$, $\mu_{V} = 0.325$, $\mu_{Sb} = -1.7$, $b = 0.5$, and $\zeta = 2$ to account for the nematic normal state.
\end{widetext}

\emph{Manifestly covariant Bloch basis (MCBB) representation.}---First of all, we rewrite the Hamiltonian $H(\boldsymbol{k})$ under the basis:
$
    \Psi^{'}_{\mathbf{k}}=\left(\psi_{Sb,\uparrow}, \psi_{V_1,\downarrow}, \psi_{V_2,\downarrow}, \psi_{V_3,\downarrow}, \psi_{Sb,\downarrow}, \psi_{V_1,\uparrow}, \psi_{V_2,\uparrow}, \psi_{V_3,\uparrow} \right)^{T} \equiv (\Psi^{'}_{\mathbf{k},\alpha_{1}}, \Psi^{'}_{\mathbf{k},\alpha_{2}})^{T}.
$
In this basis, the Hamiltonian becomes block-diagonal:
\begin{equation}
    H(\mathbf{k})=\left[\begin{array}{cc}
                        H_{\alpha_{1}\alpha_{1}}(\mathbf{k})&0\\
                        0 & H_{\alpha_{2}\alpha_{2}}(\mathbf{k})
                         \end{array}\right],
\end{equation}
where $ H_{\alpha_{1}\alpha_{1}}(\mathbf{k})=H^{*}_{\alpha_{2}\alpha_{2}}(\mathbf{k})$.

Then we diagonalize the Hamiltonian and choose a pair of degenerate orthonormal eigenstates near the Fermi surface as the target bands: $\left|\boldsymbol{k}, n_{2}, \alpha_{1}\right\rangle$, $\left|\boldsymbol{k},n_{2},\alpha_{2}\right\rangle$. The projection matrix $\Tilde{U}_{\mathbf{k}}$ is defined as:
\begin{equation}
    \Tilde{U}_{\mathbf{k}}\Psi_{\mathbf{k}}^{'}=\Tilde{\Phi}_{\mathbf{k}}=(\phi_{\mathbf{k},n_{2},\alpha_1},\phi_{\mathbf{k},n_{2},\alpha_2})^{T},
\end{equation}
where $\phi_{\mathbf{k},n_{2},\alpha_1}$ and $\phi_{\mathbf{k},n_{2},\alpha_2}$ are the annihilation operator for the two degenerate bands ($E_{n_{2},\alpha_1}(\mathbf{k})=E_{n_{2},\alpha_2}(\mathbf{k})$) near the Fermi surface. The projection of the spin operators can be expressed as: 
\begin{equation}
\begin{aligned}
    \sigma'_{x}(\mathbf{k})&=\Tilde{U}_{\mathbf{k}}(\sigma_{x}\otimes I_{4\times4} )\Tilde{U}_{\mathbf{k}}^{\dagger},\\
    \sigma'_{y}(\mathbf{k})&=\Tilde{U}_{\mathbf{k}}(\sigma_{y}\otimes diag([1,-1,-1,-1]) )\Tilde{U}_{\mathbf{k}}^{\dagger},\\
    \sigma'_{z}(\mathbf{k})&=\Tilde{U}_{\mathbf{k}}(\sigma_{z}\otimes diag([1,-1,-1,-1]) )\Tilde{U}_{\mathbf{k}}^{\dagger},\\
\end{aligned}
\end{equation}
where $I_{4\times4}$ is the four-dimensional identity matrix and $\sigma_{i}~(i=x,y,z)$ are the Pauli matrices. It is worth mentioning that because we changed the order of the basis, the spin operator is $\sigma_{i}\otimes diag([1,-1,-1,-1])$ insted of $\sigma_{i}\otimes I_{4\times 4}$ for $i=y,z$. One can find that $\sigma'_{z}(\mathbf{k})$ is naturally proportional to the Pauli matrix $\rho_{z}$:
\begin{equation}
    \sigma'_{z}(\mathbf{k})= a_{33}(\mathbf{k}) \rho_{z} \equiv \lambda_{\mathbf{k}}\rho_{z}.
\end{equation}
However, the other two spin operations are mixture of Pauli matrices $\rho_{x}$ and $\rho_{y}$:
\begin{equation}
\begin{aligned}
    \sigma'_{x}(\mathbf{k})&=a_{11}(\mathbf{k})\rho_{x}+a_{12}(\mathbf{k})\rho_{y},\\
    \sigma'_{y}(\mathbf{k})&=a_{21}(\mathbf{k})\rho_{x}+a_{22}(\mathbf{k})\rho_{y},\\
\end{aligned}
\end{equation}
The relative phase between the two eigenstates, $\left|\boldsymbol{k}, n_{2}, \alpha_{1}\right\rangle$ and $\left|\boldsymbol{k}, n_{2}, \alpha_{2}\right\rangle$, determines the form of the parameter $a_{ij}(\mathbf{k})$. To construct the MCBB, we adjust this relative phase so that the spin operators $\sigma'_{x}(\mathbf{k})$ and $\sigma'_{y}(\mathbf{k})$ transform identically to the real spin operators under the symmetry operations of the group that describes the normal-state symmetry of RV$_3$Sb$_5$. For example, in the presence of nematic states ($\zeta\neq 1$), the point group describing the normal-state symmetry of RV$_3$Sb$_5$ is $D_{2h}$, whose generators are $C_{2z}$, $C_{2x}$, and $I$. Then we can derive that the $a_{ij}(\mathbf{k})$ must satisfy the condition:
\begin{equation}
    \begin{aligned}
        a_{ij}(C_{2z}^{-1}\mathbf{k})&=a_{ij}(\mathbf{k}),\\
        a_{ij}(C_{2x}^{-1}\mathbf{k})&=(-1)^{\delta_{ij}}a_{ij}(\mathbf{k}),\\
        a_{ij}(I^{-1}\mathbf{k})&=a_{ij}(\mathbf{k}),\\
    \end{aligned}
\end{equation}
where the $\delta_{ij}$ is the Kronecker delta function.

When the external magnetic field $\mathbf{B}$ is applied, the coupling between real spin and the magnetic field can be projected onto pseudo-spin space:
\begin{equation}
    H_{coupling}(\mathbf{k})=-\mu_B \sum_{i} B_{i} \sigma'_{i}(\mathbf{k}).
\end{equation}
Thus, the effective single-band Hamiltonian in the MCBB \cite{fischer2023superconductivity} can be written as:
\begin{equation}
H_0(\boldsymbol{k})=\sum_{\boldsymbol{k} \alpha} \xi_{\boldsymbol{k}} c_{\boldsymbol{k} \alpha}^{\dagger} c_{\boldsymbol{k} \alpha}+\sum_{\boldsymbol{k}, \alpha, \beta} c_{\boldsymbol{k} \alpha}^{\dagger} (\bm{g}(\boldsymbol{k}) \cdot \bm{\rho})_{\alpha \beta} c_{\boldsymbol{k} \beta} ,
\end{equation}
where the effective magnetic field is defined as $ g_j(\boldsymbol{k})=-\mu_B \sum_i B_i a_{i j}(\boldsymbol{k})$.

\newpage
\begin{widetext}
\title {Supplementary material for  “Odd-parity topological superconductivity in kagome metal RbV$_3$Sb$_5$”}
  
\maketitle

\setcounter{figure}{0}  
\setcounter{section}{0}
\renewcommand\thefigure{S\Alph{section}\arabic{figure}}
\setcounter{equation}{0}
\renewcommand\theequation{S\arabic{equation}}

\section*{\bf{\uppercase\expandafter{SUPPLEMENTARY NOTE 1: Linearized gap equation in MCBB representation}}}
\subsection{Linearized gap equation}
The linearized gap equation in the MCBB reads
\begin{equation}
\Delta_{s_1 s_2}\left(\boldsymbol{k}\right)=-k_B T \sum_{\boldsymbol{k}^{\prime}, n, s_1^{\prime}, s_2^{\prime}, s_3^{\prime}, s_4^{\prime}} V_{s_1 s_2 s_2^{\prime} s_1^{\prime}}\left(\boldsymbol{k}, \boldsymbol{k}^{\prime}\right) G_{s_1^{\prime} s_3^{\prime}}^0\left(\boldsymbol{k}^{\prime}, i \omega_n\right) \Delta_{s_3^{\prime}, s_4^{\prime}}\left(\boldsymbol{k}^{\prime}\right) G_{s_2^{\prime} s_4^{\prime}}^0\left(-\boldsymbol{k}^{\prime},-i \omega_n\right),
\end{equation}
with the gap function
$\widehat{\Delta}(\boldsymbol{k})=\left\{\psi(\boldsymbol{k}) \hat{\rho}^0+\mathbf{d}(\boldsymbol{k}) \cdot \hat{\bm{\rho}}\right\} i \hat{\rho}^y$. The Green's function $\hat{G}^0\left(\boldsymbol{k}, i \omega_n\right)$ is obtained from the equation
\begin{equation}
\left\{\left(i \omega_n-\xi_{\boldsymbol{k}}\right) \hat{\rho}^0-\mathbf{g}_{\boldsymbol{k}} \cdot \hat{\bm{\rho}}\right\} \hat{G}^0\left(\boldsymbol{k}, i \omega_n\right)=\hat{\rho}^0,
\end{equation}
where $\omega_n=\pi k_B T(2 n+1)$ is the Fermionic Matsubara frequency. The solution to this equation is
\begin{equation}
\hat{G}^0\left(\boldsymbol{k}, i \omega_n\right)=G^{(+)}\left(\boldsymbol{k}, i \omega_n\right) \hat{\rho}^0+G^{(-)}\left(\boldsymbol{k}, i \omega_n\right) \hat{\bm{\rho}} \cdot \hat{g}_{\boldsymbol{k}},
\end{equation}
in which
\begin{equation}
G^{( \pm)}\left(\boldsymbol{k}, i \omega_n\right)=\frac{1}{2}\left\{G_{+}\left(\boldsymbol{k}, i \omega_n\right) \pm G_{-}\left(\boldsymbol{k}, i \omega_n\right)\right\} \quad \text { and } \quad G_\lambda=\frac{1}{i \omega_n-\tilde{\xi}_{\boldsymbol{k} \lambda}},
\end{equation}
with $\lambda= \pm$, $\tilde{\xi}_{\boldsymbol{k}, \lambda}=\xi_{\boldsymbol{k}}+\lambda \left|\mathbf{g}_{\boldsymbol{k}}\right|$ and $\hat{g}_{\boldsymbol{k}}=\mathbf{g}_{\boldsymbol{k}}/\left|\mathbf{g}_{\boldsymbol{k}}\right|$. We consider a general interaction:
\begin{equation}
V_{ s_1 s_2 s_3 s_4}\left(\boldsymbol{k}, \boldsymbol{k}^{\prime}\right)=\sum_a v_a\Delta_{a, s_1 s_2}\left(\boldsymbol{k}\right)\Delta^{\dagger}_{a,s_3s_4}\left(\boldsymbol{k}^{\prime}\right).
\end{equation}
Then, the linearized gap equation can be easily simplified to
 \begin{equation}
     -d/k_BT=\sum_{n \boldsymbol{k}^{\prime} a} v_{a} \operatorname{Tr} \left[\widehat{\Delta}_a\left(\boldsymbol{k}\right)\widehat{\Delta}^{-1}\left(\boldsymbol{k}\right)\right]\operatorname{Tr}\left[\hat{G}^0\left(\boldsymbol{k}^{\prime}, i \omega_n\right) \widehat{\Delta}\left(\boldsymbol{k}^{\prime}\right) \hat{G}^{0T}\left(-\boldsymbol{k}^{\prime},-i \omega_n\right) \widehat{\Delta}_a^{\dagger}\left(\boldsymbol{k}^{\prime}\right)\right],
 \end{equation}
where $d$ is the dimension of the gap function.

We then focus on the odd-parity pairing channel $\widehat{\Delta}(\boldsymbol{k})\propto \widehat{\Delta}_a(\boldsymbol{k})=\left[\mathbf{d}_a(\boldsymbol{k}) \cdot \hat{\bm{\rho}}\right] i \hat{\rho}^y$:
\begin{equation}
    \begin{aligned}
   -1 =&\sum_{n \boldsymbol{k}^{\prime}}\frac{v_a}{2\beta}  \operatorname{Tr} \left\{\left[G^{(+)}\left(\boldsymbol{k}^{\prime}, i \omega_n\right)\rho^0 +G^{(-)}\left(\boldsymbol{k}^{\prime}, i \omega_n\right) \hat{\bm{\rho}} \cdot \hat{g}_{\boldsymbol{k}^{\prime}}\right] \mathbf{d}_a(\boldsymbol{k}^{\prime}) \cdot \hat{\bm{\rho}}\right.  \\ &\left.\left[G^{(+)}\left(-\boldsymbol{k}^{\prime}, -i \omega_n\right)\rho^0 -G^{(-)}\left(-\boldsymbol{k}^{\prime}, -i \omega_n\right) \hat{\bm{\rho}} \cdot \hat{g}_{-\boldsymbol{k}^{\prime}}\right]\mathbf{d}_a^*(\boldsymbol{k}^{\prime})\cdot\hat{\bm{\rho}}  \right\}.
   \end{aligned}
\end{equation}

To further simplify the equation above, we use the identity $(\mathbf{a} \cdot \mathbf{\sigma})(\mathbf{b} \cdot \mathbf{\sigma})=(\mathbf{a} \cdot \mathbf{b}) \sigma^0+i(\mathbf{a} \times \mathbf{b}) \cdot \mathbf{\sigma}$. 
As a result, the linearized gap equation can be rewritten as follows:
\begin{equation}
-v_a^{-1}=\sum_{ \boldsymbol{k}^{\prime}}(\dots),
\end{equation}
where $(\dots)$ are the sum of the following three terms
\begin{equation}
\sum_{n}\beta^{-1}\left[G^{(+)}\left(\boldsymbol{k}^{\prime}, i \omega_n\right)G^{(+)}\left(-\boldsymbol{k}^{\prime}, -i \omega_n\right)-G^{(-)}\left(\boldsymbol{k}^{\prime}, i \omega_n\right)G^{(-)}\left(-\boldsymbol{k}^{\prime}, -i \omega_n\right)\right]|\mathbf{d}_a(\boldsymbol{k}^\prime)|^2,
\end{equation}
\begin{equation}
\sum_{n}\beta^{-1}\left[G^{(-)}\left(\boldsymbol{k}^{\prime}, i \omega_n\right)G^{(+)}\left(-\boldsymbol{k}^{\prime}, -i \omega_n\right)\hat{g}_{\boldsymbol{k}^{\prime}}+G^{(+)}\left(\boldsymbol{k}^{\prime}, i \omega_n\right)G^{(-)}\left(-\boldsymbol{k}^{\prime}, -i \omega_n\right)\hat{g}_{-\boldsymbol{k}^{\prime}}\right]\cdot\left[i\mathbf{d}_a(\boldsymbol{k}^\prime)\times\mathbf{d}_a^*(\boldsymbol{k}^{\prime})\right],
\end{equation}
\begin{equation}
\sum_{n}\beta^{-1}G^{(-)}\left(\boldsymbol{k}^{\prime}, i \omega_n\right)G^{(-)}\left(-\boldsymbol{k}^{\prime}, -i \omega_n\right)\left\{\left(\hat{g}_{\boldsymbol{k}^{\prime}}\cdot\hat{g}_{-\boldsymbol{k}^{\prime}}+1\right)|\mathbf{d}_a(\boldsymbol{k}^\prime)|^2-2\text{Re}\left[\left(\hat{g}_{\boldsymbol{k}^{\prime}}\cdot\mathbf{d}_a(\boldsymbol{k}^\prime)\right)\left(\hat{g}_{-\boldsymbol{k}^{\prime}}\cdot\mathbf{d}_a^*(\boldsymbol{k}^\prime)\right)\right]\right\}.
\end{equation}
in which $\beta=(k_BT)^{-1}$, and
\begin{equation}
\beta^{-1}\sum_{n}\left[G^{(+)}\left(\boldsymbol{k}^{\prime}, i \omega_n\right)G^{(+)}\left(-\boldsymbol{k}^{\prime}, -i \omega_n\right)-G^{(-)}\left(\boldsymbol{k}^{\prime}, i \omega_n\right)G^{(-)}\left(-\boldsymbol{k}^{\prime}, -i \omega_n\right)\right]=\frac{\sinh \beta\xi_{\boldsymbol{k}^{\prime}}}{2\xi_{\boldsymbol{k}^{\prime}} \left ( \cosh \beta| \mathbf{g}_{\boldsymbol{k}^{\prime}} |+\cosh \beta \xi_{\boldsymbol{k}^{\prime}}  \right ) },
\end{equation}
\begin{equation}
\beta^{-1}\sum_{n}\left[G^{(-)}\left(\boldsymbol{k}^{\prime}, i \omega_n\right)G^{(+)}\left(-\boldsymbol{k}^{\prime}, -i \omega_n\right)+G^{(+)}\left(\boldsymbol{k}^{\prime}, i \omega_n\right)G^{(-)}\left(-\boldsymbol{k}^{\prime}, -i \omega_n\right)\right]=\frac{- | \mathbf{g}_{\boldsymbol{k}^{\prime}} |\sinh \beta\xi_{\boldsymbol{k}^{\prime}}+\xi_{\boldsymbol{k}^{\prime}}\sinh \beta| \mathbf{g}_{\boldsymbol{k}^{\prime}} |     }{2 \left ( \xi^2_{\boldsymbol{k}^{\prime}}-| \mathbf{g}_{\boldsymbol{k}^{\prime}} |^2 \right )\left ( \cosh \beta| \mathbf{g}_{\boldsymbol{k}^{\prime}} |+\cosh \beta \xi_{\boldsymbol{k}^{\prime}}  \right )  },
\end{equation}
\begin{equation}
\beta^{-1}\sum_{n}\left[G^{(-)}\left(\boldsymbol{k}^{\prime}, i \omega_n\right)G^{(+)}\left(-\boldsymbol{k}^{\prime}, -i \omega_n\right)-G^{(+)}\left(\boldsymbol{k}^{\prime}, i \omega_n\right)G^{(-)}\left(-\boldsymbol{k}^{\prime}, -i \omega_n\right)\right]=0,
\end{equation}
\begin{equation}
\beta^{-1}\sum_{n}G^{(-)}\left(\boldsymbol{k}^{\prime}, i \omega_n\right)G^{(-)}\left(-\boldsymbol{k}^{\prime}, -i \omega_n\right)=\frac{| \mathbf{g}_{\boldsymbol{k}^{\prime}} |\left ( | \mathbf{g}_{\boldsymbol{k}^{\prime}} |\sinh \beta\xi_{\boldsymbol{k}^{\prime}}-\xi_{\boldsymbol{k}^{\prime}}\sinh \beta| \mathbf{g}_{\boldsymbol{k}^{\prime}} |   \right )  }{4\xi_{\boldsymbol{k}^{\prime}} \left ( \xi^2_{\boldsymbol{k}^{\prime}}-| \mathbf{g}_{\boldsymbol{k}^{\prime}} |^2 \right )\left ( \cosh \beta| \mathbf{g}_{\boldsymbol{k}^{\prime}} |+\cosh \beta \xi_{\boldsymbol{k}^{\prime}}  \right )  }.
\end{equation}

\subsection{Anisotropic upper critical field}\label{secupper}
In the absence of magnetic fields, the zero-field linearized gap equation is
\begin{equation}
-v_a^{-1}=\sum_{ \boldsymbol{k}^{\prime}} \frac{|\mathbf{d}_a(\boldsymbol{k}^\prime)|^2 \sinh \beta_{c0}\xi_{\boldsymbol{k}^{\prime}}}{2\xi_{\boldsymbol{k}^{\prime}} \left ( 1+\cosh \beta_{c0} \xi_{\boldsymbol{k}^{\prime}}  \right )  },
\end{equation}
$T_{c0}=(k_B\beta_{c0})^{-1}$ is the zero-field critical temperature. When there is a finite magnetic field $\mathbf{g}_{\boldsymbol{k}}=\mathbf{g}_{-\boldsymbol{k}}$ (the system has inversion symmetry), the linearized gap equation becomes
\begin{equation}
\begin{aligned}
     0=\sum_{ \boldsymbol{k}^{\prime}}&\frac{|\mathbf{d}_a(\boldsymbol{k}^\prime)|^2}{2\xi_{\boldsymbol{k}^{\prime}}  }
     \left(\frac{\xi_{\boldsymbol{k}^{\prime}}\left ( \xi_{\boldsymbol{k}^{\prime}}\sinh \beta\xi_{\boldsymbol{k}^{\prime}}-| \mathbf{g}_{\boldsymbol{k}^{\prime}} |\sinh \beta| \mathbf{g}_{\boldsymbol{k}^{\prime}} |   \right )}{\left ( \xi^2_{\boldsymbol{k}^{\prime}}-| \mathbf{g}_{\boldsymbol{k}^{\prime}} |^2 \right )\left ( \cosh \beta| \mathbf{g}_{\boldsymbol{k}^{\prime}} |+\cosh \beta \xi_{\boldsymbol{k}^{\prime}}  \right )  }-\frac{ \sinh \beta_{c0}\xi_{\boldsymbol{k}^{\prime}}}{ 1+\cosh \beta_{c0} \xi_{\boldsymbol{k}^{\prime}}    }\right)\\
     +&\frac{\xi_{\boldsymbol{k}^{\prime}}\sinh \beta| \mathbf{g}_{\boldsymbol{k}^{\prime}}|-|\mathbf{g}_{\boldsymbol{k}^{\prime}} |\sinh \beta\xi_{\boldsymbol{k}^{\prime}}     }{2 \left ( \xi^2_{\boldsymbol{k}^{\prime}}-| \mathbf{g}_{\boldsymbol{k}^{\prime}} |^2 \right )\left ( \cosh \beta| \mathbf{g}_{\boldsymbol{k}^{\prime}} |+\cosh \beta \xi_{\boldsymbol{k}^{\prime}}  \right )  }\left(\hat{\text{g}}_{\boldsymbol{k}^{\prime}}\cdot\left(i\mathbf{d}_a(\boldsymbol{k}^\prime)\times\mathbf{d}_a^*(\boldsymbol{k}^{\prime})\right)+\frac{| \mathbf{g}_{\boldsymbol{k}^{\prime}} |}{\xi_{\boldsymbol{k}^{\prime}}}|\hat{\text{g}}_{\boldsymbol{k}^{\prime}}\cdot\mathbf{d}_a(\boldsymbol{k}^\prime)|^2\right).
\end{aligned}
\end{equation}
With the transition temperature at the zero field case, we can calculate the decoupling constant; then we can solve the gap equation under a finite magnetic field. During this calculation, the parameters are: $t=1$, $t_2=0.5$, $\mu_{V}=0.325$, $\mu_{Sb}=-1.7$, $b=0.5$, $\zeta=2$, $T_{c0}=0.02$.

\section*{\bf{\uppercase\expandafter{SUPPLEMENTARY NOTE 2: Topological superconductivity}}}
 
\subsection{superconducting order parameters in D$_{2h}$ point group}
  Considering the nematic normal state, the point group describing the symmetry of the normal states of RbV$_{3}$Sb$_{5}$ is D$_{2h}$. 
  The superconducting gap functions, which are the eigenstates of the linearized gap equation, form the bases for the irreducible representations (irreps) of the point group $D_{2h}$. Each irrep represents one individual superconducting channel. The irreps of $D_{2h}$ are shown in Table.II.
  \begin{table}[h]\label{d2h}
      \centering
      \begin{tabular}{cccc}
      \hline \hline  
           irreps & $C_{2z}$ & $C_{2x}$ & $I$\\ \hline
           $A_{g}$ & 1        & 1        &  1  \\ \hline
           $B_{1g}$ & 1        & -1        &  1  \\ \hline
           $B_{2g}$ & -1        & -1        &  1  \\ \hline
           $B_{3g}$ & -1        & 1        &  1  \\ \hline
           $A_{u}$ & 1        & 1        &  -1  \\ \hline
           $B_{1u}$ & 1        & -1        &  -1  \\ \hline
           $B_{2u}$ & -1        & -1        &  -1  \\ \hline
           $B_{3u}$ & -1        & 1        &  -1  \\ \hline
      \end{tabular}
      \caption{The generators and irreps of the $D_{2h}$ point group.}
  \end{table}
  
Here, since we are considering odd parity superconducting states that are nodal and exhibit in-plane magnetic moments, only the superconducting order parameters corresponding to irreps with odd parity are listed in Table.~\ref{REP}.
\begin{table}[h]
     \caption{The irreducible representations (irreps) of $D_{2h}$ point group and the corresponding order parameters. For our purposes, only the irreps with odd parity are listed. In this table, $k_{1}=k_{x}$, $k_2=k_x/2+\sqrt{3}k_y/2$, and $k_{3}=k_{2}-k_{1}$. In addition, $\mathbf{\hat{x}}$, $\mathbf{\hat{y}}$, and $\mathbf{\hat{z}}$ represent axial-vectors oriented along the $x$, $y$, and $z$ directions, respectively.}
    \centering
    \begin{ruledtabular}
    \begin{tabular}{cccc}\label{REP}
    Representations & order parameter $\mathbf{d}(\mathbf{k})$ \\ \hline
    $A_{u}$ & $sin(k_{1})\mathbf{\hat{x}}$; $(sin(k_{2})+sin(k_{3}))\mathbf{\hat{y}}$.\\ \hline
    $B_{1u}$ & $(sin(k_{2})+sin(k_{3}))\mathbf{\hat{x}}$; $sin(k_{1})\mathbf{\hat{y}}$. \\
    \hline
    $B_{2u}$ & $(sin(k_{2})+sin(k_{3}))\mathbf{\hat{z}}$\\ \hline
    $B_{3u}$ & $sin(k_{1})\mathbf{\hat{z}}$ \\ 
    \end{tabular}
    \end{ruledtabular}
    \label{REP1}
\end{table}
To obtain the nonunitary pairing states, we select the superconducting order parameter by combining order parameters belonging to different irreps of the point group $D_{2h}$, under the assumption that they have nearly degenerate transition temperatures, as discussed in the main text.

\subsection{Topological properties of superconducting state}
As discussed in previous sections and the main text, the nodal superconducting order parameters with nonzero in-plane magnetic moments can be constructed by combining order parameters from different irreps. The superconducting order parameter can generally be expressed as:
\begin{equation}\label{dv}
    \mathbf{d}=\Delta_{0} sin(k_{1})(cos(\theta)cos(\phi),sin(\theta)cos(\phi),i sin(\phi)).
\end{equation}
The parameters $\theta$ and $\phi$ depend on the relative amplitudes between superconducting order parameters belonging to different irreps. We choose $\theta=0$ without loss of generality. In addition, the intrinsic polarization of the pairing state is determined by the decoupling coefficient and can be characterized by the parameter $\phi$. The highest transition temperature was chosen to be the state with $\phi=0.3\pi$. When considering the superconducting order parameter shown in Eq.(\ref{dv}), the total BdG Hamiltonian can be rewritten as:
\begin{equation}\label{hbdg}
    H_{BdG}(\mathbf{k})=\left(\begin{array}{cc}H_{0}(\mathbf{k})&0\\0&-H_{0}^{*}(-\mathbf{k})\end{array}\right)-\delta_{1}sin(k_{1})\tau_{1}\otimes\sigma_{3}-\delta_{2}sin(k_{1})\tau_{2}\otimes\sigma_{0}-\delta_{3}sin(k_{1})\tau_{2}\otimes\sigma_{1},
\end{equation}
where we rewrite the d-vector as: $\mathbf{d}=(\delta_{1},\delta_{2},i\delta_{3})$, $\tau_{i}$, and $\sigma_{i}$ with $i=0,1,2,3$ being the identity matrix ($0$), and the Pauli matrices ($1,2,3$). The BdG Hamiltonian preserves the effective time-reversal symmetry: $T'=i\tau_{0}\otimes \sigma_{1}K$ ($T'^{2}=1$) and particle-hole symmetry: $C=\tau_{1}\otimes \sigma_{0}K$ ($C^2=1$), thus belonging to the BDI class. The time reversal symmetry and particle-hole symmetry are defined by:
\begin{equation}
    \begin{aligned}
        T'H(\mathbf{k})T'^{-1}=H(-\mathbf{k}),\\
        CH(\mathbf{k})C^{-1}=-H(-\mathbf{k}),
    \end{aligned}
\end{equation}
respectively.
\begin{figure}[ht]
    \centering
    \includegraphics[width=0.5\textwidth]{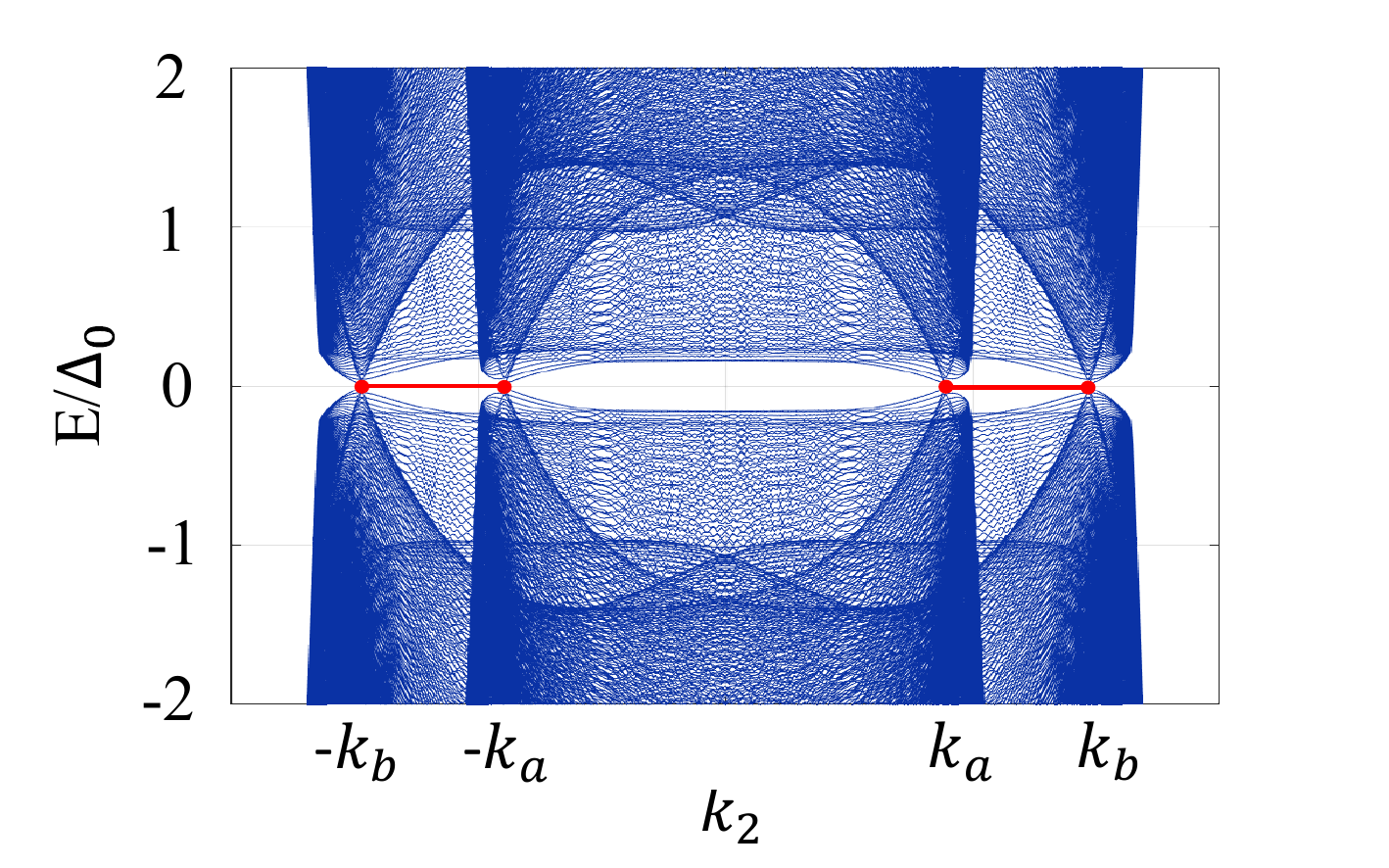}
    \caption{The energy spectrum of $H_{BdG}(\mathbf{k})$ with open boundary condition in the $k_1$ direction and periodic boundary condition in the $k_2$ direction. During this calculation, the amplitude of the gap $\Delta_{0}$ is $0.03 t$, and the number of lattices along the x direction is $n=1000$.}
    \label{zfcopband}
\end{figure}
We calculate the energy spectrum of $H_{BdG}(\mathbf{k})$ with open boundary conditions in the $k_1$ direction and periodic boundary conditions in the $k_2$ direction; there are Majorana zero modes between the nodes of the superconducting spectrum, as shown in Fig.~\ref{zfcopband}. The coordinates of these nodes are $(0,-k_{b}),~(0,-k_{a}),~(0,k_{a}),~(0,k_{b})$.

To calculate the winding number of $H_{BdG}$, we first check whether this Hamiltonian can be decoupled into several diagonal blocks.
For $H_{BdG}(\mathbf{k})$ shown in Eq.(\ref{hbdg}), we can find a unitary transformation matrix $U_{TH}$ that can make the Hamiltonian block-diagonal.
\begin{align}
    U_{TH}=\frac{1}{\sqrt{2}}\left(\begin{array}{cccc}
      -ie^{-i\zeta}&e^{i\zeta}&0&0\\
      0&0&ie^{i\zeta}&e^{-i\zeta}\\
      ie^{-i\zeta}&e^{i\zeta}&0&0\\
      0&0&-ie^{i\zeta}&e^{-i\zeta}
    \end{array}\right),
\end{align}
where $\zeta=\frac{\pi}{4}+\frac{1}{2}arctan(\frac{\delta_{1}}{\delta_{2}})$.

The block-diagonal Hamiltonian can be expressed as:
\begin{equation}
    U_{TH}H_{BdG}(\mathbf{k})U_{TH}^{-1}=\left(\begin{array}{cc}
        H_{1}(\mathbf{k}) & 0 \\
         0 & H_{2}(\mathbf{k})
    \end{array}\right),
\end{equation}
where these two blocks are:
\begin{equation}
    H_{1}(\mathbf{k})=\left( \begin{array}{cc}
       H_{0}(\mathbf{k})  & (\delta_{3}-\sqrt{\delta_{1}^2+\delta_{2}^2})sin(k_{x}) \\
       (\delta_{3}-\sqrt{\delta_{1}^2+\delta_{2}^2})sin(k_{x})  & -H_{0}^{*}(-\mathbf{k})
    \end{array} \right)
\end{equation}
and
\begin{equation}
     H_{2}(\mathbf{k})=\left( \begin{array}{cc}
       H_{0}(\mathbf{k})  & -(\delta_{3}+\sqrt{\delta_{1}^2+\delta_{2}^2})sin(k_{x}) \\
       -(\delta_{3}+\sqrt{\delta_{1}^2+\delta_{2}^2})sin(k_{x})  & -H_{0}^{*}(-\mathbf{k})
    \end{array} \right).
\end{equation}
Then, for $H_{BdG}$, two chiral symmetry operators can be found:
\begin{equation}
    \begin{aligned}
        S_{1}&=-\tau_{1}\otimes\sigma_{1},\\
        S_{2}&=-cos(\lambda) \tau_{1}\otimes\sigma_{0}+sin(\lambda) \tau_{2}\otimes \sigma_{3},
    \end{aligned}
\end{equation}
where $\lambda=arctan(\frac{\delta_{1}}{\delta_{2}})$.
These two chiral symmetry operators can also be transformed into block-diagonal by $U_{TH}$:
\begin{equation}\label{diagchiral}
\begin{aligned}
      U_{TH}^{\dagger}S_{1}U_{TH}=-\tau_{3}\otimes\sigma_{2}\\
      U_{TH}^{\dagger}S_{2}U_{TH}=-\tau_{0}\otimes\sigma_{2}.
\end{aligned}
\end{equation}
Therefore, for these two blocks, the chiral symmetry is both $S=\sigma_{2}$. The winding number for each Hamiltonian block at a certain $k_{2}=k_{2,0}$ can be calculated by the equation:
\begin{equation}
    W(k_{2,0})=\frac{i}{4\pi}\int_{-\pi}^{\pi}d k_{1} Tr(S H(k_{1},k_{2,0})^{-1}\partial_{k_{1}}H(k_{1},k_{2,0})).
\end{equation}
The winding number of $H_{1}$ and $H_{2}$ between point nodes can be obtained ($\phi=0.3 \pi$):
\begin{equation}
\begin{aligned}
         W_{1}(k_{2})=\left(\begin{array}{lc}
          -1 & for\    \ k_{a}<|k_{2}|<k_{b} \\
          0 & for\     \ other\     \ case
    \end{array}\right)
\end{aligned}
\end{equation}
and
\begin{equation}
    W_{2}(k_{2})=\left(\begin{array}{lc}
          1 & for\    \ k_{a}<|k_{2}|<k_{b} \\
          0 & for\     \ other \      \ case
    \end{array}\right).
\end{equation}
respectively. These two Hamiltonian blocks do not couple to each other in the zero-field case.

\subsection{Method of tunneling spectra calculation}

The tunneling conductance between the tunneling lead and the boundary of the kagome superconductor is calculated using the scattering matrix approach:
\begin{equation}\label{tunc}
\begin{gathered}
    G_{\boldsymbol{c}}(E, T)= \int d E^{\prime}\left(-\frac{\partial f\left(E^{\prime}, T\right)}{\partial E^{\prime}}\right)G_{\boldsymbol{c}}(E^{\prime}, 0),\\
G_{\boldsymbol{c}}(E, 0)=\frac{e^2}{h} \operatorname{Tr} \left[I-r_{e e}^{\dagger}\left(E\right) r_{e e}\left(E\right)+r_{h e}^{\dagger}\left(E\right) r_{h e}\left(E\right)\right],
\end{gathered}
\end{equation}
where $f(E, T)$ is the Fermi distribution function. An electron coming from the tunneling lead will be scattered when it tunnels into the kagome superconductor from the boundary. The reflection matrices in Eq.~(\ref{tunc}) can be understood as follows: for an incoming electron, there is a $\left|r_{e e}\right|^2$ chance of being scattered back as an electron and a $\left|r_{h e}\right|^2$ chance of being scattered back as a hole. Near zero bias, the Andreev reflection process will be resonant for the edge Majorana flat band. Technically, the reflection matrix can be calculated using the recursive Green’s function method:
\begin{equation}
r_{a b}(E)=-I \delta_{a b}+i \tilde{\Gamma}_a^{1 / 2}(E) G_{n n}^R(E) \tilde{\Gamma}_b^{1 / 2}(E) .
\end{equation}
Here, $a, b \in\{e, h\}$, the broadening function $\tilde{\Gamma}(E)=i\left(\Sigma(E)-\Sigma^{\dagger}(E)\right)$, where $\Sigma(E)$ is the self energy of the semi-infinite tunneling lead. $G^R(E)=1 /(E+i \eta-H)$ is the retarded Green's function. $n$ labels the position in the tunneling lead.

\end{widetext}

\end{document}